\documentclass[%
 reprint,
 amsmath,amssymb,
 aps,
 showpacs, showkeys,
]{revtex4-2}

\usepackage{graphicx}
\usepackage{dcolumn}
\usepackage{bm}

\begin{document}

\title{Linearization Principle: The Geometric Origin of Nonlinear Fokker-Planck Equations}

\author{Hiroki Suyari}
 \email{suyari@faculty.chiba-u.jp}
 \email{suyarilab@gmail.com}
 \thanks{ORCID: 0000-0003-2624-0902}
\affiliation{%
 Graduate School of Informatics, Chiba University, 1-33, Yayoi-cho, Inage-ku,
Chiba 263-8522, Japan\\
}

\date{May 21, 2026}

\begin{abstract}
Anomalous diffusion and power-law distributions are observed in various complex systems.
To provide a consistent dynamical foundation for these phenomena, we present a geometric derivation of the nonlinear Fokker-Planck equation by introducing the Linearization Principle directly at the dynamical stage.
By identifying the generalized chemical potential as the natural dynamical ansatz, we construct a general thermodynamic framework where the drift term remains linear in the probability density, preserving the standard form of the Einstein relation.
Within this framework, we show that the $q$-deformed geometry, corresponding to Tsallis statistics, exhibits a fundamental duality between the dynamic index $q$ and the thermodynamic index $2-q$: the stationary state is a $q$-Gaussian distribution that minimizes a free energy functional defined by a generalized entropy of index $2-q$.
We prove the $H$-theorem for the derived equation and demonstrate its application to the harmonic oscillator and the free particle.
This framework describes anomalous diffusion without relying on ad-hoc constraints or phenomenological nonlinear drift forces.
\end{abstract}

                            
\maketitle


\section{Introduction}
\label{sec:introduction}

The dynamical origin of anomalous diffusion and power-law distributions remains a central problem in statistical physics.
While the static properties of these ubiquitous phenomena are often well-captured by generalized entropies, such as the Tsallis entropy~\cite{Tsallis1988}, their corresponding dynamical foundations---specifically, the exact form of the Fokker-Planck equation (FPE) that generates these statistics---have remained a subject of debate~\cite{Plastino1995, Frank2005, Curado1991}.

Standard approaches to derive a nonlinear FPE for power-law statistics typically encounter a physical dilemma.
To ensure that the stationary solution matches the observed $q$-Gaussian distributions, these models often necessitate state-dependent diffusion coefficients or phenomenological nonlinear drift terms.
A prevalent formulation relies on ``escort distributions''~\cite{Beck1993, Tsallis2009}, which act as auxiliary measures to maintain mathematical consistency.
However, the introduction of nonlinear drift terms---forces that depend on the probability density itself---violates the independence of particles from the external potential.
Furthermore, relying on auxiliary distributions obscures the physical meaning of the observables.

In this paper, we resolve this dilemma by introducing the \textit{Linearization Principle} at the dynamical stage.
Rather than modifying the thermodynamics at the static level to fit the observed statistics, we ground the dynamics in a generalized chemical potential as the fundamental ansatz.
By adopting the $\phi$-logarithm as the generalized coordinate for the thermodynamic driving force, we formulate a nonlinear FPE that preserves the linearity of the drift term and the standard form of the Einstein relation.
In particular, we focus on the $q$-logarithm of Tsallis statistics as a representative case to demonstrate how the geometric structure dictates the macroscopic dynamics.
This dynamical derivation uncovers a fundamental duality: while the local evolution is governed by the index $q$, the global thermodynamic stability is ensured by a Lyapunov functional based on an entropy of index $2-q$.
This approach eliminates the need for ad-hoc constraints or escort distributions.

The theory yields $q$-Gaussian stationary states in harmonic potentials and recovers the correct scaling laws for anomalous diffusion in free space.
Furthermore, this classical geometric framework serves as the foundation for the macroscopic states of 1D quantum fluids detailed in our companion paper~\cite{SuyariPartV}, bridging nonlinear statistical mechanics and strongly correlated quantum systems.

The paper is organized as follows.
In Sec.~\ref{sec:derivation}, we derive the nonlinear FPE from the geometric properties of the $q$-logarithm.
In Sec.~\ref{sec:stationary}, we discuss the stationary states and the generalized Einstein relation.
In Sec.~\ref{sec:entropy}, we prove the $H$-theorem and detail the $q \leftrightarrow 2-q$ duality.
Finally, in Sec.~\ref{sec:applications}, we apply the theory to the harmonic oscillator and the free particle.


\section{Geometric Derivation of Nonlinear Dynamics}
\label{sec:derivation}

Rather than postulating a specific entropy functional \textit{a priori} to fit a desired stationary state, we introduce the \textit{Linearization Principle} directly into the dynamical stage. This geometric principle uniquely determines the macroscopic nonlinear diffusion from a generalized thermodynamic driving force, addressing the dynamical origin of power-law distributions.

\subsection{The Linearization Principle via $\phi$-deformed Geometry}
In the standard linear response theory of non-equilibrium thermodynamics, the probability flux $J(x,t)$ is driven by the gradient of the chemical potential $\mu$:
\begin{equation}
    J = - \frac{p}{\gamma} \nabla \mu,
    \label{eq:standard_flux}
\end{equation}
where $\gamma$ is the friction coefficient. In the Boltzmann-Gibbs framework, the chemical potential is given by $\mu = k_B T \ln p + V(x)$, leading to the standard linear Fokker-Planck equation (FPE).

To extend this to a broader class of complex systems without introducing phenomenological ad-hoc drift forces, we employ the generalized $\phi$-logarithm introduced by Naudts \cite{Naudts2011}. For a strictly positive, non-decreasing function $\phi(u)$, the $\phi$-logarithm is defined as:
\begin{equation}
    \ln_\phi(p) := \int_{1}^{p} \frac{1}{\phi(v)} dv.
\end{equation}
The derivative satisfies $\frac{d}{dp}\ln_\phi(p) = \frac{1}{\phi(p)}$, which characterizes the underlying geometry of the state space.

The \textit{Linearization Principle} postulates that the thermodynamic driving force remains linear with respect to the generalized geometric coordinate $\ln_\phi(p)$. Accordingly, we introduce the generalized chemical potential as a dynamical ansatz:
\begin{equation}
    \mu_\phi := k_B T \ln_\phi(p) + V(x).
    \label{eq:mu_phi}
\end{equation}
Substituting Eq.~(\ref{eq:mu_phi}) into the flux Eq.~(\ref{eq:standard_flux}), we obtain:
\begin{align}
    J &= - \frac{p}{\gamma} \nabla (k_B T \ln_\phi(p) + V) \nonumber \\
      &= - \frac{k_B T}{\gamma} \frac{p}{\phi(p)} \nabla p - \frac{p}{\gamma} \nabla V.
\end{align}
By applying the standard Einstein relation $D = k_B T / \gamma$, the generalized flux is expressed as:
\begin{equation}
    J = - D \frac{p}{\phi(p)} \nabla p - \frac{1}{\gamma} p \nabla V. 
    \label{eq:phi_flux}
\end{equation}
Equation (\ref{eq:phi_flux}) demonstrates that the Linearization Principle is not a mere change of variables tautologically derived from a specific probability distribution. Rather, it acts as a fundamental constructive principle: specifying the geometric structure of the thermodynamic space ($\phi(p)$) uniquely determines the density-dependent effective nonlinear diffusion coefficient $\mathcal{D}(p) = D \frac{p}{\phi(p)}$, while strictly preserving the linear structure of the external drift term. While we focus on the $q$-deformed geometry in the following sections, Eq. (\ref{eq:phi_flux}) implies that choosing any valid geometric structure $\phi(p)$ constructively yields its corresponding macroscopic nonlinear diffusion equation. This demonstrates the genuine generality of the Linearization Principle beyond the specific framework of the $q$-logarithm.

\subsection{Specialization to the $q$-logarithm and the Porous Medium Equation}
Having established the general mechanism, we focus on the specific geometric choice $\phi(p) = p^q$ ($q \in \mathbb{R}$), which characterizes scale-invariant interactions. This choice recovers the Tsallis $q$-logarithm:
\begin{equation}
    \ln_q(p) := \frac{p^{1-q} - 1}{1-q} \quad (q \neq 1).
\end{equation}
Under this specific geometry, the density-dependent diffusion coefficient in Eq.~(\ref{eq:phi_flux}) takes the precise form of a power law:
\begin{equation}
    \mathcal{D}(p) = D \frac{p}{p^q} = D p^{1-q}.
\end{equation}
Substituting the resulting flux $J = - D p^{1-q} \nabla p - \frac{1}{\gamma} p \nabla V$ into the continuity equation $\frac{\partial p}{\partial t} + \nabla \cdot J = 0$ yields the nonlinear Fokker-Planck equation (NFPE):
\begin{equation}
    \frac{\partial p}{\partial t} = \nabla \cdot \left( D p^{1-q} \nabla p + \frac{1}{\gamma} p \nabla V \right).
    \label{eq:NFPE}
\end{equation}
This equation is equivalent to the porous medium equation with an external drift, often associated with the dynamics of $q$-statistics. While the final form coincides with existing results \cite{Plastino1995, TsallisBukman1996}, our derivation clarifies that the power-law nonlinearity arises naturally from the requirement of a linear driving force within a $q$-deformed geometric manifold. 

It is crucial to emphasize the fundamental difference in the stage at which the respective \textit{ansätze} are introduced. In standard nonextensive thermostatistics, the $q$-deformed ansatz is typically introduced at the static level---specifically, by maximizing a generalized entropy to fit a (quasi-)equilibrium steady-state distribution, while the corresponding dynamical evolution often requires ad-hoc phenomenological forces. In contrast, our approach introduces the geometric ansatz directly at the dynamical stage via the generalized chemical potential. This formulation fundamentally dictates the dynamical evolution of the statistical quantities, naturally yielding the stationary power-law distributions as a dynamical consequence rather than imposing them as an \textit{a priori} static requirement. As discussed in Section \ref{sec:entropy}, this specific choice of $\phi(p)=p^q$ leads to a remarkable duality between the dynamical index $q$ and the thermodynamic index $2-q$.


\section{Stationary Solutions and Equilibrium Properties}
\label{sec:stationary}

\subsection{Stationary State Condition}
The stationary state $p_{st}(x)$ is characterized by the vanishing of the probability flux, $J = 0$.
From Eq.~(\ref{eq:phi_flux}) with $\phi(p)=p^q$, this condition yields:
\begin{equation}
    - D p_{st}^{1-q} \nabla p_{st} - \frac{1}{\gamma} p_{st} \nabla V(x) = 0.
\end{equation}
Dividing by $p_{st}$, we obtain:
\begin{equation}
    D p_{st}^{-q} \nabla p_{st} = - \frac{1}{\gamma} \nabla V(x).
\label{eq:stationary_condition}
\end{equation}
Using the identity $p_{st}^{-q} \nabla p_{st} = \nabla (\ln_q p_{st})$, Eq.~(\ref{eq:stationary_condition}) is expressed in terms of the natural coordinate:
\begin{equation}
    D \nabla (\ln_q p_{st}) = - \frac{1}{\gamma} \nabla V(x).
\end{equation}

\subsection{The $q$-Gaussian Distribution}
Integration with respect to $x$ yields:
\begin{equation}
    \ln_q p_{st}(x) = - \frac{1}{\gamma D} V(x) + C,
\end{equation}
where $C$ is an integration constant.
Applying the inverse $q$-logarithm, $\exp_q(u) := [1 + (1-q)u]^{1/(1-q)}$, yields the stationary distribution:
\begin{equation}
    p_{st}(x) = \frac{1}{Z_q} \exp_q \left( - \beta_q V(x) \right), \label{eq:q_gaussian}
\end{equation}
where $Z_q$ is the normalization constant and $\beta_q$ is the inverse temperature:
\begin{equation}
    \beta_q := \frac{1}{\gamma D} = \frac{1}{k_B T}.
\end{equation}
For a harmonic potential $V(x) = \frac{1}{2} k x^2$, Eq.~(\ref{eq:q_gaussian}) reproduces the $q$-Gaussian distribution, a typical characteristic of systems exhibiting anomalous diffusion.

\subsection{Generalized Einstein Relation}
\label{sec:einstein}
In many previous phenomenological extensions of the Fokker-Planck equation, the relationship between the diffusion coefficient and temperature required complex redefinitions or the introduction of $q$-dependent factors.
In contrast, the present geometric formulation leaves the standard Einstein relation $D = k_B T / \gamma$ preserved.
By identifying $\ln_q p$ as the natural coordinate for the chemical potential, the fundamental link between fluctuation and dissipation retains its classical linear form, even as it generates highly nonlinear stationary states.


\section{Entropy and Free Energy: The Dual Nature}
\label{sec:entropy}

The stability of the stationary solution and the irreversibility of time evolution are governed by a Lyapunov functional, identified as the generalized free energy of the system.
The geometric derivation of the NFPE necessitates a generalized entropy governed by a dual index $2-q$.

\subsection{Free Energy Functional}
Define the generalized free energy functional $\mathcal{F}[p]$ as the sum of the external potential energy and a generalized entropic term:
\begin{equation}
    \mathcal{F}[p] := \int V(x) p(x) dx + k_B T \int \Psi(p) dx,
\end{equation}
where $\Psi(p)$ ensures the NFPE constitutes a gradient flow of $\mathcal{F}$.
The variation of the free energy with respect to $p$ is:
\begin{equation}
    \frac{\delta \mathcal{F}}{\delta p} = V(x) + k_B T \Psi'(p).
\end{equation}
Comparing this with the generalized chemical potential $\mu_q = V(x) + k_B T \ln_q p$, we identify:
\begin{equation}
    \Psi'(p) = \ln_q p.
\end{equation}

\subsection{The $2-q$ Entropy and $H$-theorem}
Integrating $\Psi'(p) = \ln_q p$ with the condition $\Psi(0)=0$ yields:
\begin{equation}
    \Psi(p) = \int \frac{p^{1-q}-1}{1-q} dp = \frac{1}{1-q} \left( \frac{p^{2-q}}{2-q} - p \right).
\end{equation}
This functional generates the Tsallis entropy of index $2-q$:
\begin{equation}
    S_{2-q}[p] := \frac{1 - \int p^{2-q} dx}{(2-q) - 1}.
\end{equation}
This geometric structure yields the $H$-theorem.
The time derivative of the free energy is:
\begin{equation}
    \frac{d\mathcal{F}}{dt} = \int \frac{\delta \mathcal{F}}{\delta p} \frac{\partial p}{\partial t} dx = - \int \gamma p \left| v \right|^2 dx \leq 0,
\end{equation}
where $v = - \frac{1}{\gamma} \nabla \mu_q$ is the velocity field.
This inequality ensures that the system monotonically approaches the stationary state, minimizing the free energy associated with the index $2-q$.

\textit{Boundary conditions for the $H$-theorem.}---The proof of macroscopic irreversibility ($d\mathcal{F}/dt \le 0$) relies on integration by parts.
This operation is mathematically justified only when the probability density $p(x,t)$ and the associated probability current $J(x,t)$ vanish at the boundaries ($|x| \to \infty$).
For compact distributions ($q < 1$), this condition is trivially satisfied at the physical edges of the finite support.
For heavy-tailed $q$-Gaussian distributions ($q > 1$), this requirement imposes a specific constraint on the asymptotic behavior of the potential $V(x)$ to ensure the boundary terms vanish, thereby guaranteeing the validity of the generalized $H$-theorem.

\subsection{Duality of Indices}
This establishes a duality: a non-equilibrium system whose dynamics is governed by the index $q$ possesses a thermodynamic structure governed by the dual index $2-q$.
This duality eliminates the necessity of escort distributions, restoring the use of standard expectation values for physical observables.
As demonstrated by Wada and Scarfone~\cite{WadaScarfone2005}, formulating thermodynamics via the dual index $2-q$ preserves the standard Legendre transform structure without introducing auxiliary probability measures.


\section{Applications to Specific Systems}
\label{sec:applications}

We apply the derived NFPE to two fundamental systems: the harmonic oscillator and the free particle.

\subsection{The Harmonic Oscillator: $q$-Gaussian Statistics}
Consider a harmonic potential, $V(x) = \frac{1}{2} k x^2$, with spring constant $k > 0$.
Equation~(\ref{eq:q_gaussian}) yields the stationary distribution:
\begin{equation}
\begin{split}
    p_{st}(x) &= \frac{1}{Z_q} \exp_q \left( - \frac{\beta_q k}{2} x^2 \right) \\
              &= \frac{1}{Z_q} \left[ 1 - (1-q) \frac{\beta_q k}{2} x^2 \right]^{\frac{1}{1-q}}.
\end{split}
\end{equation}
This is the $q$-Gaussian distribution.
Unlike the standard Gaussian ($q=1$), the $q$-Gaussian exhibits power-law tails for $q > 1$, characterized by the asymptotic behavior $p(x) \sim |x|^{-2/(q-1)}$.
For $q < 1$, the distribution has compact support, vanishing beyond a finite cutoff.
This confirms that the geometric derivation recovers the $q$-Gaussian as the unique stationary state under a linear restoring force.
The $q$-logarithmic structure thereby connects the deterministic nonlinear dynamics directly with the stationary statistics.

\begin{figure}[t]
    \centering
    \includegraphics[width=\linewidth]{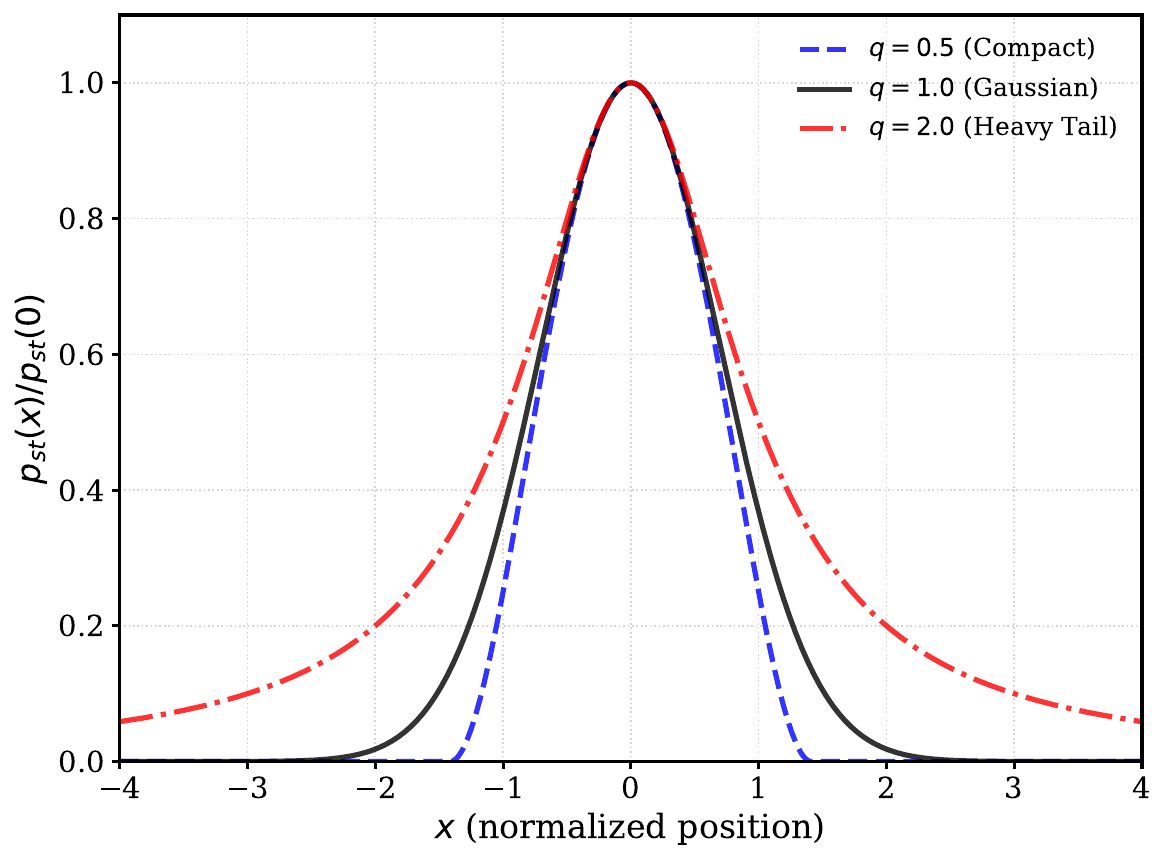}
    \caption{Comparison of stationary distributions $p_{st}(x)$ for a particle in a harmonic potential.
The curves are normalized to unity at $x=0$. The solid black line represents the standard Gaussian distribution ($q=1$).
The dot-dashed red line shows a heavy-tailed $q$-Gaussian ($q=2.0$), typical for super-diffusive systems.
The dashed blue line shows a $q$-Gaussian with compact support ($q=0.5$), typical for sub-diffusive systems.}
    \label{fig:q_gaussian}
\end{figure}

\subsection{The Free Particle: Anomalous Diffusion}
For a free particle ($V=0$), the NFPE (Eq.~\ref{eq:NFPE}) reduces to the porous medium equation:
\begin{equation}
    \frac{\partial p}{\partial t} = D \frac{\partial}{\partial x} \left( p^{1-q} \frac{\partial p}{\partial x} \right).
\label{eq:nonlinear_diffusion}
\end{equation}
Applying the self-similar scaling ansatz $p(x,t) = t^{-\alpha} F(x/t^{\alpha})$, where $F$ is the scaling function representing the invariant profile, yields the dimensional constraint matching the time derivative ($t^{-(\alpha+1)}$) and the diffusion term ($t^{\alpha q - 4\alpha}$):
\begin{equation}
    -(\alpha + 1) = \alpha q - 4\alpha.
\end{equation}
The scaling exponent is determined as:
\begin{equation}
    \alpha = \frac{1}{3-q}.
\end{equation}
Consequently, the mean square displacement $\langle x^2(t) \rangle \propto t^{2\alpha}$ obeys the anomalous diffusion law:
\begin{equation}
    \langle x^2(t) \rangle \propto t^{\frac{2}{3-q}}.
\end{equation}
The parameter $q$ classifies the diffusion regime:
\begin{itemize}
    \item $q = 1$: normal diffusion ($\langle x^2 \rangle \propto t$).
    \item $q < 1$: sub-diffusion, typical in porous media.
    \item $1 < q < 3$: super-diffusion, observed in turbulent systems or active matter.
\end{itemize}
This establishes a geometric link between the parameter $q$ and the macroscopic anomalous diffusion exponent.


\section{Concluding Remarks}
\label{sec:conclusion}

In this work, we presented a geometric derivation of the nonlinear Fokker-Planck equation, grounded in the \textit{Linearization Principle}.
By recognizing that the Linearization Principle, when applied to the dynamical evolution, dictates the natural coordinate system for the thermodynamic driving force, we formulated a kinetic theory that preserves the linear structure of the drift term and the standard Einstein relation.
This clarifies that the nonlinear diffusion arises naturally from a fundamental dynamical ansatz rather than being imposed as an \textit{a priori} static requirement.

Importantly, this geometric framework is not restricted to the specific case of Tsallis statistics.
By employing the generalized $\phi$-logarithm, we demonstrated that specifying any valid geometric structure for the state space constructively yields its corresponding macroscopic nonlinear diffusion equation.
This establishes the genuine generality of the Linearization Principle for nonequilibrium complex systems.

Within this general framework, focusing on the $q$-deformed geometry yields a central result: the duality between the dynamic index $q$ and the thermodynamic index $2-q$.
We showed that while the local trajectories are governed by $q$-arithmetic, the global stability of the stationary state is determined by a Lyapunov functional based on the Tsallis entropy of index $2-q$.
This duality clarifies the choice of entropic indices in non-extensive statistical mechanics and eliminates the need for auxiliary concepts such as escort distributions.

Applied to the harmonic oscillator and the free particle, the theory reproduces the $q$-Gaussian distributions observed in confined systems and yields the scaling laws of anomalous diffusion in free space.

Finally, we note the mathematical origin of the Linearization Principle.
While the theory was presented primarily using the language of nonlinear dynamics, the specific natural coordinate $\ln_q p$ identified here possesses a precise geometric meaning.
In the context of Information Geometry~\cite{Amari2016}, this coordinate corresponds to the \textit{dual affine coordinate} of a dually flat manifold associated with the $q$-exponential family~\cite{AmariOhara2011, OharaWada2010}.
This indicates that the nonlinear diffusion process can be geometrically understood as a linear gradient flow projected onto a curved statistical manifold.
This geometric perspective connects the classical anomalous diffusion discussed here with the macroscopic states of 1D quantum fluids discussed in our companion paper~\cite{SuyariPartV}, revealing a link between information geometry and the dynamics of strongly correlated systems.

Future work will extend this framework to multi-dimensional systems.
The present geometric formulation provides a consistent basis for analyzing complex non-equilibrium phenomena.

\begin{acknowledgments}
The author thanks the anonymous referees for their constructive comments and suggestions, which have helped to improve the manuscript.
\end{acknowledgments}


\bibliography{apssamp}

\end{document}